# Ultrawideband solid-state terahertz phase shifter electrically modulated by tunable conductive interface in total internal reflection geometry

Xudong Liu[1*], Daosong Yu[1], Chuanfu Sun[1], Zhijie Mei[1], Hao Chen[1], Jianbin Xu[2*], Yiwen Sun[1*]

[1] Department of Biomedical Engineering, School of Medicine, Shenzhen University, Shenzhen 518060, China

[2] Department of Electronic Engineering, The Chinese University of Hong Kong, Hong Kong SAR, China

* xdliu@szu.edu.cn, jbxu@ee.cuhk.edu.hk, ywsun@szu.edu.cn

**Abstract**: Phase modulation plays a crucial role in various terahertz applications, including biomedical imaging, high-rate communication, and radar detection. Existing terahertz phase shifters typically rely on tuning the resonant effect of metamaterial structures to achieve a narrow bandwidth phase shift. However, the terahertz band offers a wide bandwidth resource, which has great advantages in high longitudinal resolution detection, high-capacity communication, spectral imaging and so on. Here, we propose and demonstrate an ultrawideband terahertz phase shifting mechanism that utilizes an optical conductivity tuneable interface combined with a non-resonant metasurface operating in the total internal reflection geometry. This approach effectively modulates the phase of the reflected terahertz signal in an ultrawideband. To implement this mechanism, we designed a structure consisting of graphene-loaded non-resonant periodic metal microslits arranged in the total internal reflection geometry. By controlling the gate voltage of the graphene within a range of $\pm 5$ *V*, an averaged ~120° continuous phase shift in the frequency range of 0.4 to 1.2 THz was achieved. Notably, in the frequency range of 1 to 1.2 THz, the phase modulation exhibited a linear relationship with the driving voltage. Our device demonstrated minimal fluctuations in the reflected amplitude, with a deviation of less than 1 dB and an insertion loss of less than 10 dB. Additionally, the modulation speed of this solid-state device reached the kHz level. Remarkably, the phase modulation bandwidth ($\Delta f/f$) achieved approximately 100% of the arithmetic centre frequency at 0.8 THz, surpassing the definition of ultrawideband, which typically encompasses 20% of the centre frequency. To the best of our knowledge, this is the first and most wideband phase shifter developed for the terahertz regime to date.

**Keywords**: terahertz, ultrawideband, phase modulation, graphene

## 1. Introduction

Terahertz (THz) technology has immense potential across various fields due to its abundant bandwidth resources. Applications such as biomedical imaging [1-4], high-rate communication [5-7], material characterization [8-10], and radar detection [11, 12] can greatly benefit from advancements in THz technology. While significant progress has been made in fundamental THz devices, such as sources, detectors, and waveguides, as reviewed in 2017 [13] and 2023 [14], the development of terahertz phase modulators, especially wideband phase modulators, remains a challenging task in the THz regime. Existing techniques primarily rely on metamaterial structures that manipulate their resonant behavior between inductive and capacitive modes [15-18] to induce a phase change. For example, H.T. Chen reported the first solid-state terahertz phase modulator using a metasurface integrated GaAs Schottky diode, achieving a phase shift of approximately π/6 radians [15]. Subsequent designs have explored photoinduced vanadium dioxide coupled nanostructure [16], gate-controlled graphene metasurfaces [18], and GaN high electron mobility transistor (HEMT) [17]. However, these metamaterial-based designs are limited to narrow bandwidth operation. Although a graphene-based Brewster angle device demonstrated wideband phase modulation capability, it only achieved a discontinuous phase shift [19]. The utilization of liquid crystal material in the total internal reflection (TIR) geometry has been explored to alter the phase of reflected terahertz light [20]. However, the birefringence of liquid crystals is typically low in the terahertz region [21], resulting in limited phase shift. Another approach involved a THz polarization converter that employed ion-gel gated two graphene layers in the TIR geometry to induce phase changes for both *s*- and *p*- polarizations, but the ion-gel device operates at a slow speed and achieves a phase shift of less than 80° [22]. In this study, we propose an ultrawideband phase shifting mechanism utilizing an optically thin conductive layer in the total internal reflection (TIR) geometry at the interface (referred to as CI-TIR). This approach enables efficient phase modulation of the reflected light while preserving the wideband characteristics for both *s*- and *p*- polarizations. For this purpose, we employ a single layer of graphene as the conductive interface. Graphene is chosen for its atomic thickness, tunable sheet conductivity, and its nearly frequency-independent conductivity within the terahertz band [23, 24]. While previous work has explored the use of graphene for amplitude modulation of terahertz radiation [25-27], its potential for phase modulation has not been extensively investigated. In this paper, we demonstrate an effective phase modulation technique by utilizing a graphene-loaded non-resonant metasurface composed of a periodic arrangement of metal microslits. This structure achieves a continuous phase shift of approximately 120° in the frequency range of 0.4 to 1.2 THz. Notably, the phase modulation bandwidth ($\Delta f/f$) reaches approximately 100% at *f*= 0.8 THz, surpassing the definition of ultrawideband (UWB) which typically covers 20% of the center frequency Particularly, in the frequency range between 1 and 1.2 THz, the phase shift exhibits a linear dependence on the driving voltage. The amplitude fluctuation during phase modulation remains below 1 dB across the entire operational bandwidth, while incurring an insertion loss of approximately 10 dB. Moreover, the phase modulation speed of our device operates at the kHz level. Overall, our device showcases ultrawideband, linear, low-loss, and near-ideal phase-only modulation capabilities.

## 2. Theory and simulation

When light propagates from a dense medium to a less dense medium at an angle greater than the critical angle, total internal reflection occurs, causing the incident light to be reflected back into the dense medium. However, the incident light actually penetrates into the less dense medium as an evanescent wave and travels along the interface for a short distance. This phenomenon is accompanied by a non-zero imaginary part in the reflection coefficient, resulting in a phase shift relative to the incident light. According to Fresnel's equations, this phase shift is constant and wideband when the materials involved are non-dispersive. To manipulate the phase shift, one approach is to modify the refractive index of either the dense or less dense medium [20], but the tunable range is limited (see supplementary information). In this article, we propose a new mechanism, which uses a conductive interface in the TIR geometry (Fig. 1) to effectively tune the phase of the reflected terahertz light in a wideband. We study the phase shift in the

CI-TIR geometry from the modified Fresnel equations in Ref. [26] (Eq. 1 and 2),

$$r_s = \frac{n_1 \cos\theta_i - i \cdot \sqrt{n_1^2 \sin^2\theta_i - n_2^2} - Z_0 \sigma_s}{n_1 \cos\theta_i + i \cdot \sqrt{n_1^2 \sin^2\theta_i - n_2^2} + Z_0 \sigma_s}, \quad (1)$$

$$r_p = \frac{i \cdot n_1 \cdot \sqrt{n_1^2 \sin^2\theta_i - n_2^2} - n_2^2 \cos\theta_i - i \cdot Z_0 \sigma_s \cos\theta_i \cdot \sqrt{n_1^2 \sin^2\theta_i - n_2^2}}{i \cdot n_1 \cdot \sqrt{n_1^2 \sin^2\theta_i - n_2^2} + n_2^2 \cos\theta_i + i \cdot Z_0 \sigma_s \cos\theta_i \cdot \sqrt{n_1^2 \sin^2\theta_i - n_2^2}}, \quad (2)$$

where $n_1$ is the refractive index of dense media, $n_2$ is the refractive index of less dense media, $\theta_i$ is the supercritical incident angle, $Z_0$ is the vacuum impedance (377 Ω), and $\sigma_s$ is the optical sheet conductivity. In both *s* and *p* polarizations, considering linear, homogeneous, isotropic, and non-magnetic media, the reflection coefficient exhibits an imaginary part due to the supercritical incident angle. This imaginary part affects the phase of the reflection coefficient and can be adjusted by changing $\sigma_s$. To comprehensively analyze the impact of various parameters on the phase described by Eq. 1 and 2, a series of calculations were performed. The incident angle was varied from the critical angle to the grazing angle, while the sheet conductivity was adjusted from 0 to 10 *mS*. Additionally, six combinations of different materials were selected. The theoretical results for both *s* and *p* polarizations are illustrated in Fig. 2 and 3, respectively.

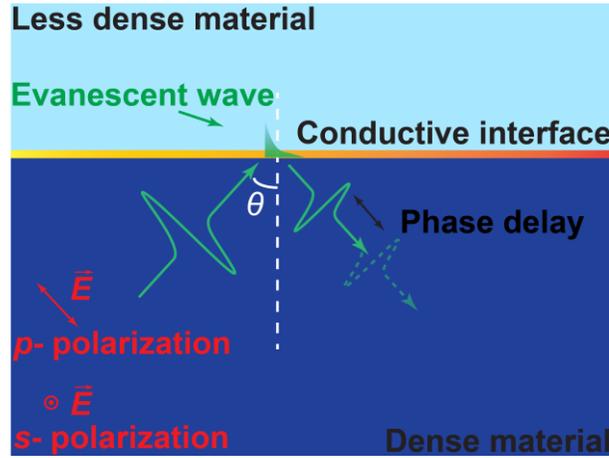

**Fig. 1. Schematic of the conductive interface total internal reflection geometry.** The lighter bule and darker bule rectangular represent less dense and dense material, respectively. The yellow line represents the conductive interface. The green arrow curves represent the incident and reflected light, which has a supercritical incident angle ($\theta$). Due to the evanescent wave in the total internal reflection, there is a phase delay in the reflected light (dashed green curve). The red arrows represent the electric field in *p*- polarization (in-plane) and *s*- polarization (out-of-plane).

The theoretical results demonstrate that the phase of the reflected light can be adjusted by manipulating the sheet conductivity in both *s* and *p* polarizations. Since the conductive interface is assumed to be isotropic, the phase shift in the CI-TIR model is wideband. Additionally, the theoretical analysis reveals that the phase shift in *s* polarization is more responsive to changes in the sheet conductivity compared to *p* polarization. This discrepancy can be attributed to the fact that *s* polarization has a greater in-plane electric field component interacting with the conductive interface than *p* polarization.

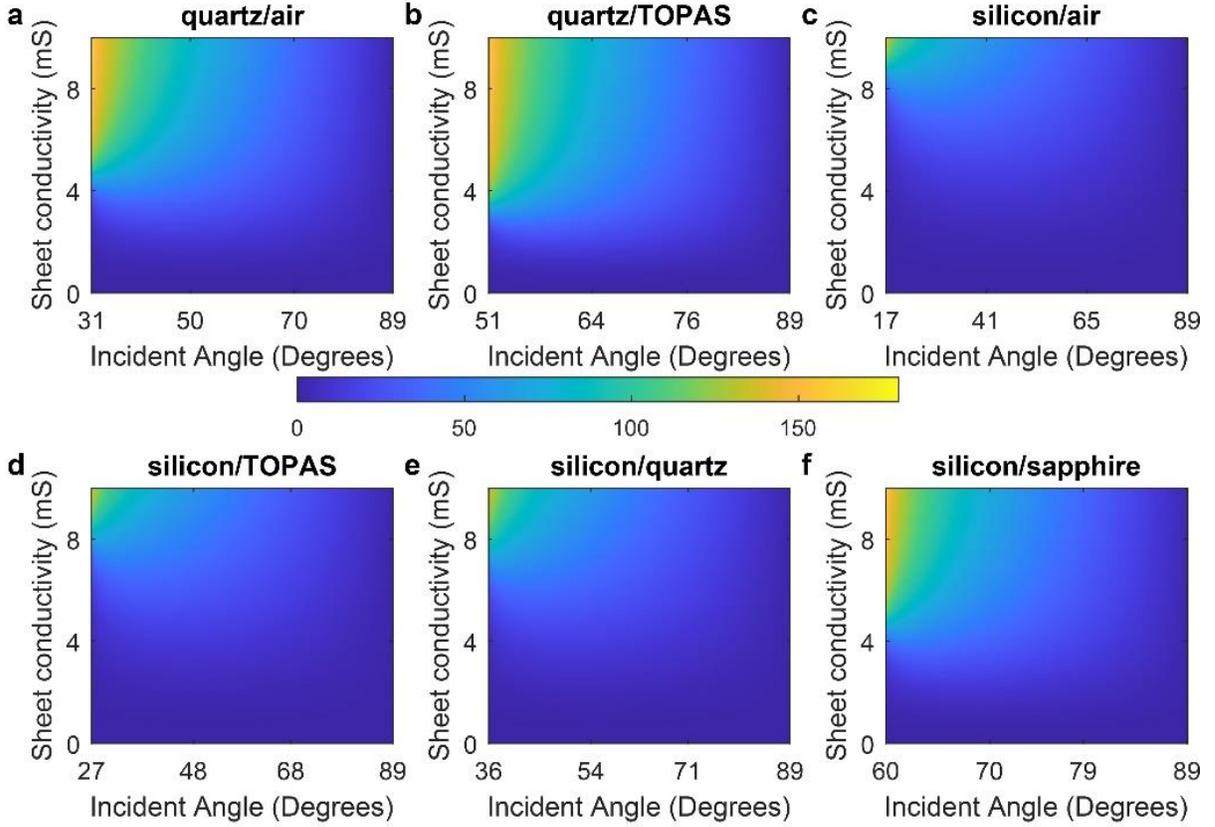

**Fig. 2. a-f, The theoretical results of phase change along with incident angle and sheet conductivity for six combinations of different materials in *s* polarization.** The incident angle is from critical angle to grazing incident, and the sheet conductivity is from 0 to 10 *mS*. The colorbar in the center represents the phase shift of the reflected light. The refractive index of the materials in calculation at terahertz band are $n_{\text{air}} \cong 1$, $n_{\text{silicon}} \cong 3.42$, $n_{\text{quartz}} \cong 2.1$, $n_{\text{TOPAS}} \cong 1.5$, and $n_{\text{saphhire}} \cong 3$.

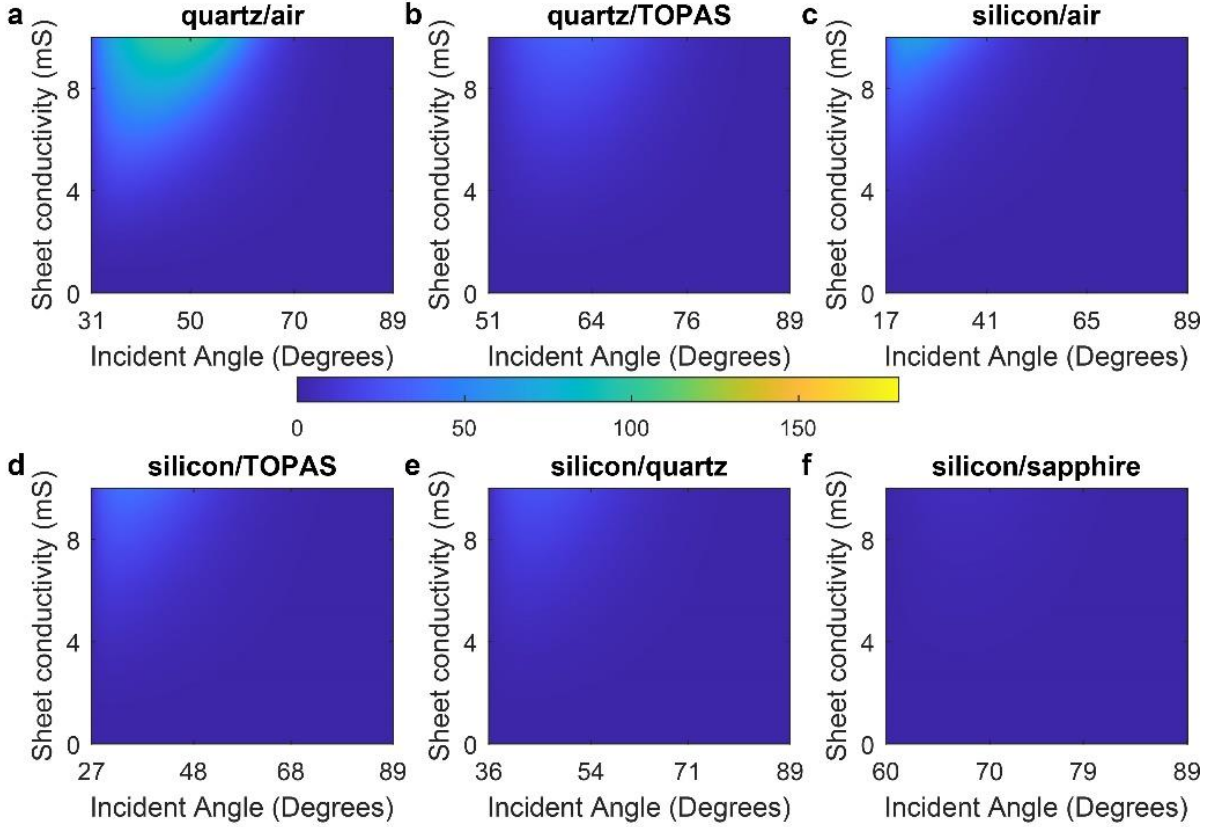

**Fig. 3. a-f, The theoretical results of phase change along with incident angle and sheet conductivity for six combinations of different materials in *p* polarization.** The incident angle is from critical angle to grazing incident, and the sheet conductivity is from 0 to 10 *mS*. The colorbar in the center represents the phase shift of the reflected light. The refractive index of the materials in calculation at terahertz band are $n_{air} \cong 1$, $n_{silicon} \cong 3.42$, $n_{quartz} \cong 2.1$, $n_{TOPAS} \cong 1.5$, and $n_{sapphire} \cong 3$.

In this article, high-resistivity silicon is chosen as the dense material due to its low-loss properties in the terahertz band. To achieve an approximately 120° phase shift at an incident angle of 30° in a silicon/air combination, the sheet conductivity of the interface needs to be adjusted within the range of 0 to approximately 8 mS (as shown in Fig. 2c). However, achieving such a high sheet conductivity electrically through a solid-state device in experimental setups can be challenging. To address this issue and reduce the required sheet conductivity while maintaining the operational bandwidth, we utilize a non-resonant electric field enhancement metasurface known as subwavelength periodic metal microslits [28-30]. This metasurface is employed to enhance the electric field intensity of the evanescent wave in the TIR model. The enhancement factor, denoted as $\eta$, is defined as the ratio of the period of the microslits (*P*) to the gap width (*g*). By optimizing this enhancement factor, we can effectively enhance the electric field intensity of the evanescent wave, enabling lower sheet conductivities to achieve the desired phase shift. The reflection coefficient with the enhancement factor can be written as:

$$r_s = \frac{n_1 \cos\theta_i - i \cdot \sqrt{n_1^2 \sin^2\theta_i - n_2^2} - \eta \cdot Z_0 \sigma_s}{n_1 \cos\theta_i + i \cdot \sqrt{n_1^2 \sin^2\theta_i - n_2^2} + \eta \cdot Z_0 \sigma_s} \quad . \tag{3}$$

The phase information in Eq. 3 has not been previously investigated and is derived here (see supplementary for details):

$$\varphi = \tan^{-1} \frac{\sqrt{n_1^2 \sin^2\theta_i - n_2^2}}{n_1 \cos\theta_i - \eta \cdot Z_0 \sigma_s} + \tan^{-1} \frac{\sqrt{n_1^2 \sin^2\theta_i - n_2^2}}{n_1 \cos\theta_i + \eta \cdot Z_0 \sigma_s} \quad . \tag{4}$$

The introduction of subwavelength metal microslits allows for a reduction in the required sheet conductivity to achieve an amplitude and phase modulation by a factor of $1/\eta$ compared to the initial value but preserves the operational bandwidth. Figure 4 illustrates the phase and intensity changes in a silicon/air model with metal microslits of various enhancement factors at an incident angle of 30° in the CI-TIR geometry. The relative phase shift and intensity are referenced to the results obtained from the TIR model without a conductive interface. Fig. 4a shows that as the enhancement factor ($\eta$) increases, the phase curve exhibits a steeper gradient. The most significant phase change occurs between 0 and 2 $mS$ sheet conductivity for an enhancement factor range of $\eta=8$ to 20. the goal is to achieve a pure phase shift while maintaining a constant intensity. However, achieving this ideal scenario is challenging in practice. Instead, the focus is on minimizing intensity fluctuations to approach a near-ideal phase shift. Fig. 4b demonstrates the corresponding intensity changes with sheet conductivity. Initially, the intensity decreases as the sheet conductivity increases from zero, reaching a minimum at a specific sheet conductivity ($\sigma_m$), after which it gradually recovers. The phase exhibits the most rapid change when the sheet conductivity is approximately $\sigma_m$. The phenomenon can be explained as follows: when the sheet conductivity is zero in the TIR model, the interface behaves as a dielectric, and the incident light is totally reflected. This reflection results in a constant phase change without any intensity attenuation. On the other hand, when the sheet conductivity approaches infinity, the interface acts as a "perfect electric conductor," leading to total reflection with a 180° phase change. In this case, the reflected intensity remains at unity as well. As the sheet conductivity varies between zero and infinity, the reflected intensity becomes less than unity and reaches its minimum at $\sigma_m$. Prior to reaching $\sigma_m$, the conductive layer exhibits a more absorptive effect, causing the reflected intensity to continuously decrease. Beyond $\sigma_m$, the conductive layer demonstrates a more reflective effect, leading to a recovery of the reflected intensity. The $\sigma_m$ point serves as a transition point between absorption and reflection, often referred to as the "absorption-to-reflection" (A-R) point. Around this transition point, the phase exhibits the most significant and drastic changes. This behavior is analogous to the inductance-capacitance resonant theory observed in metamaterials, where the inductive and capacitive behavior undergoes a transition at the resonant frequency. In the CI-TIR model with metal microslits, the phase shift is not based on resonant effects, allowing for a wideband phase shift. The range of sheet conductivity that can be achieved using a solid-state electrical graphene device is from 0.5 to 1.5 *mS* [26, 31], as indicated by the grey rectangular region in Fig. 4. Within this range, the theoretical phase changes for designs without slits, $\eta=2$, and $\eta=3$ are relatively small. However, for designs with higher enhancement factors, such as $\eta=8$, 10, 15, and 20, larger phase shifts are obtained. Specifically, the phase changes for $\eta=8$, 10, 15, and 20 are 102°, 103°, 78°, and 46°, respectively. The phase change increases as $\eta$ increases from 2 to 10. However, as $\eta$ further increases from 15 to 20, the phase change starts to decrease. This behavior is attributed to the "A-R" point, which occurs outside the range highlighted by the grey rectangular region. For designs with $\eta=8$ and 10, the reflected intensity initially decreases, then gradually increases, and remains relatively stable during the phase shift period. It is important to note that for designs with $\eta=2$ and 3, as the sheet conductivity increases from 0.5 to 2 *mS*, the reflected intensity decreases. On the other hand, for designs with $\eta=15$ and 20, the reflected intensity increases, indicating the presence of an "A-R" transition in these cases.

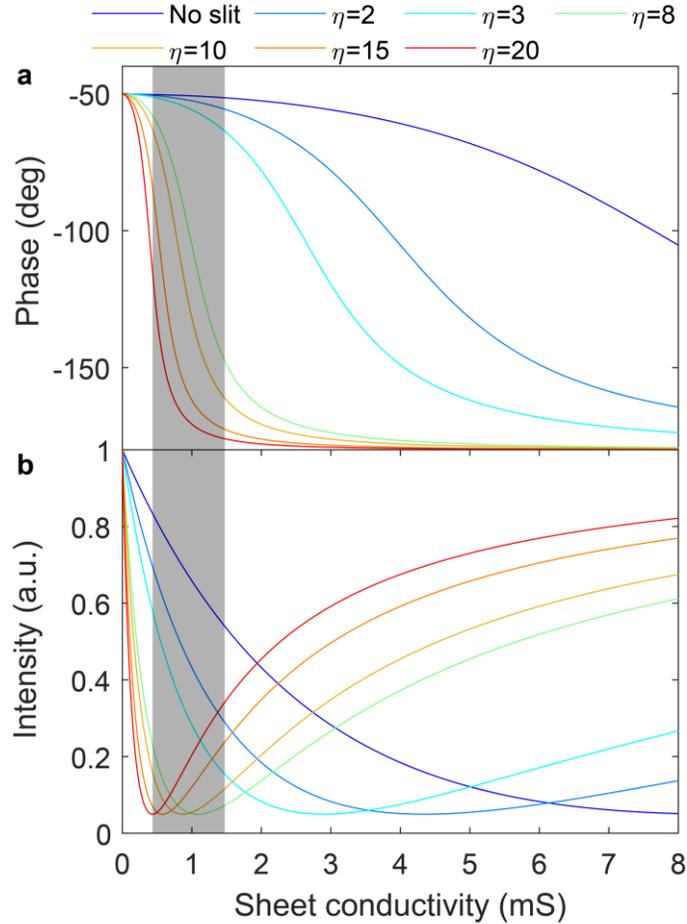

**Fig.4. The function of phase shift with sheet conductivity and different terahertz field enhancement factors ($\eta$) in the CI-TIR geometry.** The design with six enhancement factors and no microslits are chosen for calculation. The grey rectangular represents the tunable range of sheet conductivity. **a** and **b**, The phase change and reflected intensity of the reflected light as a function of optical sheet conductivity. The relative phase shift and intensity are referred to the results of TIR model without conductive interface.

## 3. Experimental methods

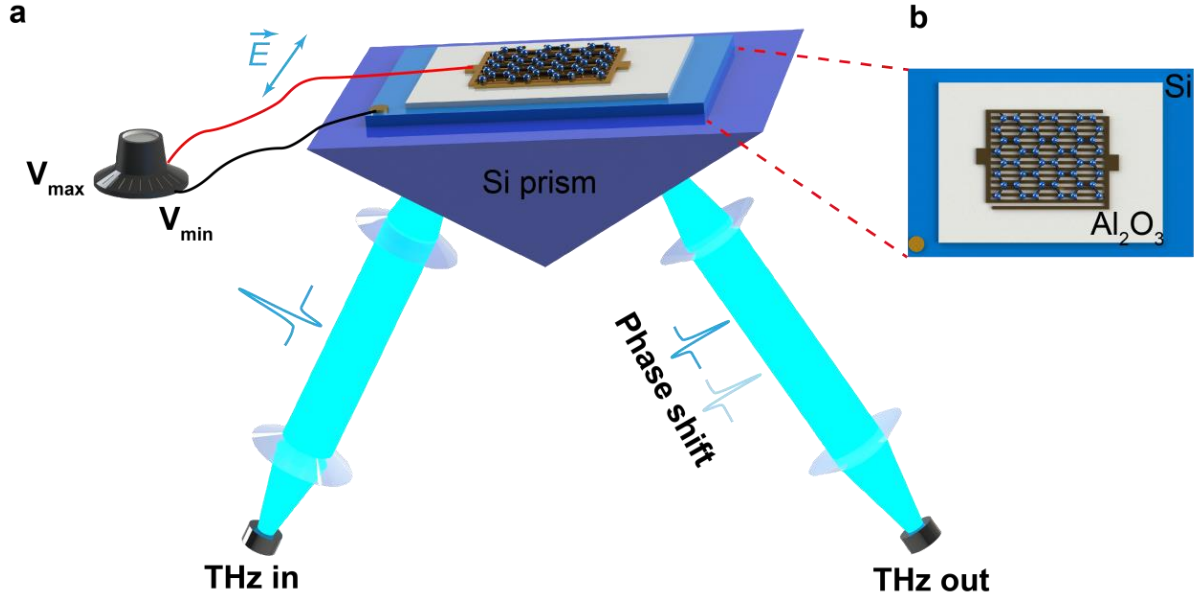

**Fig. 5. Diagram of the graphene-loaded metal microslits device in the total internal reflection geometry. a.** 3D schematic of the device in the total internal reflection geometry. A 120° isosceles triangle shaped high-resistivity silicon prism introduces a 30° supercritical incident angle to the terahertz beam. The terahertz light is *s*-polarized (out-of-paper). By applying an external bias voltage on the graphene layer, the reflected terahertz signal can be shifted in the time-domain. **b**, Schematic of the $Al_2O_3$/Si gated graphene-loaded metal microslits device.

The structure of the graphene-based device used in the experiment is depicted in Fig. 5a. It consists of a high-*κ* material $Al_2O_3$ layer, which is 10 *nm* thick and grown by atomic layer deposition (ALD), serving as the insulation layer. The $Al_2O_3$ layer is deposited on a high-resistivity silicon substrate ($R_\square$ > 6 kΩ·cm, double-side polished). On the $Al_2O_3$/Si substrate, a golden microslits pattern with dimensions of 5×5 *mm* is fabricated using standard lift-off photolithography and metallization techniques. The metal microslits have a fixed period of 20 *µm*, while the slit width varies from 2 to 10 *µm*, denoted as D2 to D10 in the paper. After the metallization process, a piece of CVD-grown graphene (5×5 *mm*) was transferred onto the metal microslits pattern. After the metallization process, a single-layer graphene, grown through chemical vapor deposition (CVD), is transferred onto the metal microslits pattern. The quality of the graphene layer is verified using Raman spectroscopy to ensure its single-layer nature (more details can be found in the supplementary information). To control the conductivity of graphene, a gated voltage is applied across the graphene layer. The voltage is swept from –5 to 5 *V* in steps of 1 *V*, taking care not to exceed the breakdown voltage of the $Al_2O_3$ insulation layer. For the experimental measurement of phase shift, a commercial terahertz time-domain spectroscopy (THz-TDs) system from TERA-K15 (Menlo Systems Inc.) is utilized. A high-resistivity silicon prism with isosceles triangular shape, where each base angle is set to 30°, is employed to provide the necessary supercritical incident angle. The incident terahertz light used in the experiment is *s*-polarized, meaning it is perpendicular to the plane of the metal microslits (out-of-paper direction in Fig. 5a). The THz-TDs system is also used to measure the sheet conductivity of graphene in the transmission geometry using a device without metal microslits. The measured results show that the graphene layer is *p*-doped, and its sheet conductivity can be tuned from approximately 0.3 *mS* (at –5 *V*) to 1.5 *mS* (at 5 *V*) in the frequency range of 0.4 to 1.2 THz (see supplementary information).

## 4. Results and discussions

The experimental results of the devices (D2 to D10) in the TIR geometry are presented in Fig. 6, 7, and 8. To analyze the amplitude and phase of the reflected terahertz waveforms, the fast Fourier transform (FFT) is employed to convert the time-domain waveform data into the frequency domain. In Fig. 6a, the peak-to-peak value of the time-domain waveform for D2 increases as the voltage changes from –5 V (0.3 *mS*) to 5 *V* (1.5 *mS*). This indicates that the graphene layer exhibits a more reflective effect, and its A-R point is left outside the conductivity tunable region. Conversely, for devices D4 to D10, the peak-to-peak values of the time-domain waveforms show an opposite trend, suggesting that the graphene layer exhibits a more absorptive effect, and the A-R points of these devices are located right outside the conductivity tunable region. These experimental observations are consistent with the theoretical predictions shown in Fig. 4. The time-domain waveform of D3 stands out as it exhibits a small variation in its peak-to-peak value and a noticeable time delay, as shown in Fig. 6b. This indicates a significant phase shift in the frequency domain. The relative phase change under various voltages, denoted as $\Delta\varphi(\omega) = |\varphi_{V(\omega)} - \varphi_{5V(\omega)}|$, is calculated with reference to the phase at 5 *V*. For D3, this relative phase change ranges over 100° from 0.4 to 1.2 THz, with an average value of approximately 120° (Fig. 6h). The theoretical phase shift for D3 (with $\eta \approx 7$) between 0.3 and 1.5 *mS* is predicted to be 102°, which aligns well with the experimental result. The phase shift gradually decreases from D3 to D10 due to the decline in the enhancement factor $\eta$. It is worth noting that the phase shift in all devices exhibits a slight increase with increasing frequency. This can be attributed to the slight frequency-dependent conductivity of graphene. In the terahertz band, the imaginary part of $\sigma$ increases slightly with frequency, while the real part decreases with frequency [19]. According to Eq. 4, this leads to a larger phase shift as the frequency increases. According to Eq. 4, this leads to a larger phase shift as the frequency increases.

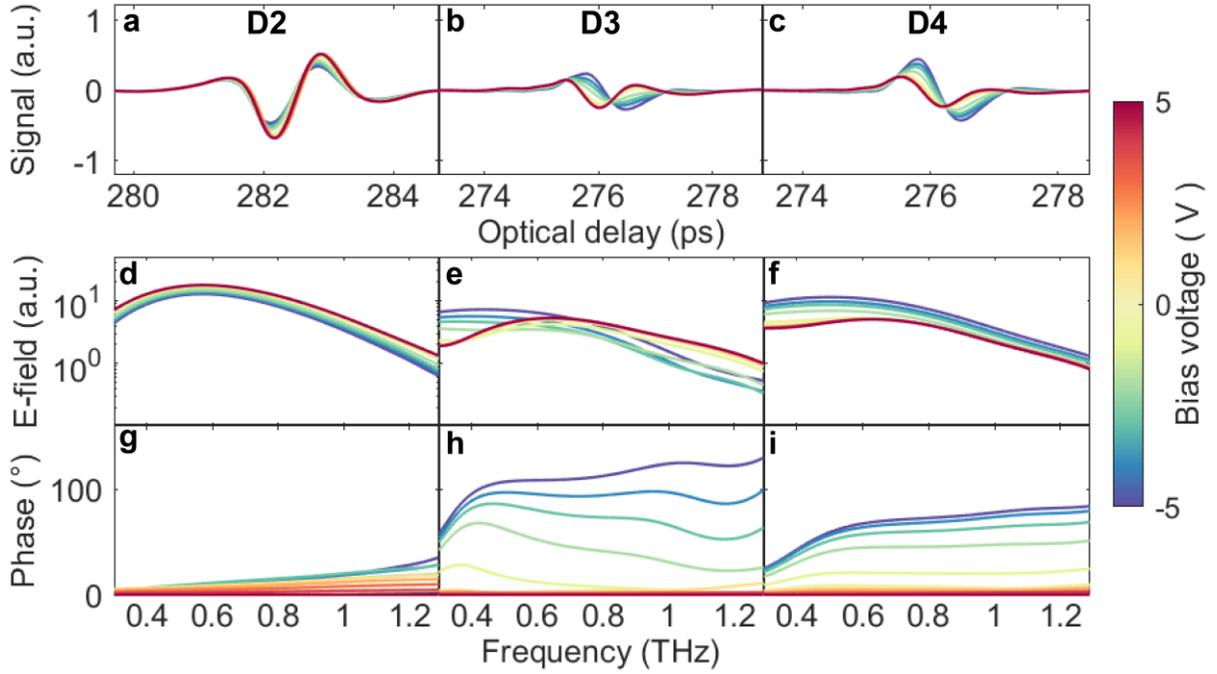

**Fig. 6. The experimental results of D2, D3, and D4 in the TIR geometry.** The reflected terahertz waveforms in time-domain (**a-c**) and frequency-domain (**d-f**) at different gate voltages. (**h-j**) The calculated corresponding relative phase shift in the reflected terahertz signal compared with the phase at 5 *V*.

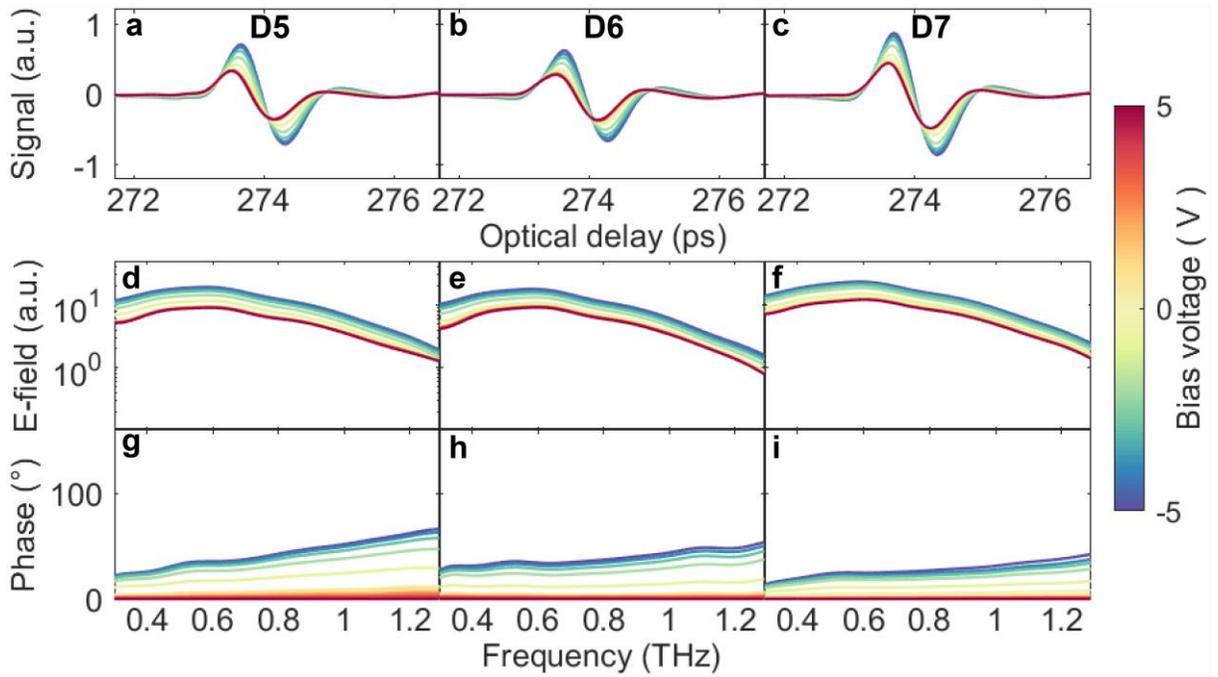

**Fig. 7. The experimental results of D5 to D7 in the TIR geometry.** The reflected terahertz waveforms in time-domain (**a-c**) and frequency-domain (**d-f**) at different gate voltages. (**h-j**) The calculated corresponding relative phase shift in the reflected terahertz signal compared with the phase at 5 *V*.

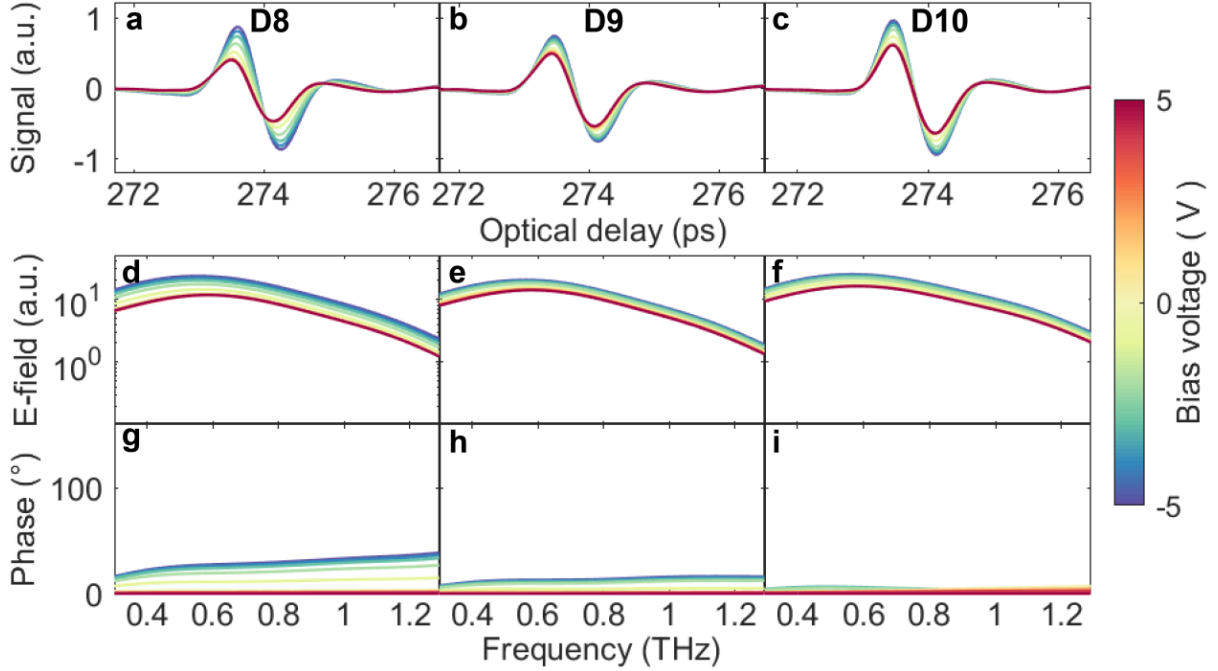

**Fig. 8. The experimental results of D8 to D10 in the TIR geometry.** The reflected terahertz waveforms in time-domain (**a-c**) and frequency-domain (**d-f**) under different gate voltages. (**h-j**) The calculated corresponding relative phase shift in the reflected terahertz signal compared with the phase at 5 *V*.

To demonstrate the continuous phase shift of device D3, a finer voltage sweeping experiment was conducted with a step size of 0.2 *V*. The terahertz time-domain waveform exhibits a gradual delay of approximately 330 *fs* as the driving voltage is swept from 5 *V* to –5 *V* (refer to supplementary information). Fig. 9a presents the relative reflected intensity of D3 compared to the bare Si prism (without the device) in the frequency range of 0.4 to 1.2 THz. The insertion loss of our design arises from two sources: 1. the entrance and exit surfaces of the Si prism, and 2. the conductive interface. The insertion loss caused by the Si prism, in this study, is approximately 5 dB, which can be mitigated by implementing anti-reflection structures on the entrance and exit surfaces [32-34]. Therefore, the insertion loss of D3 reported in this article primarily accounts for the loss from the active layer, typically less than 10 dB (Fig. 9a). This is significantly lower than the insertion losses observed in metamaterial-based narrowband phase modulators. For an ideal phase shifter, it is desirable to minimize the intensity fluctuation while shifting the phase. The reflected intensity fluctuation of D3 is less than 2 dB compared to its average value across the frequency range of 0.4 to 1.2 THz. Fig. 9b illustrates the voltage dependence of the phase shift calculated from 0.4 to 1.2 THz. From 1 to 1.2 THz, there is a quasi-linear relationship between the voltage and phase shift in the range of –5 to 0 *V*. At other frequencies, the phase shift exhibits a monotonic relationship with the voltage. The inset in Fig. 9b provides an example at *f*= 1.2 THz, demonstrating the linearity of the phase shift as a function of the driving voltage from –5 to 0 *V*. During the 130° linear phase shift process, the reflected intensity fluctuation remains within ±2 dB compared to its mean value. This can be compensated by adjusting the source output power or incorporating a wideband amplitude modulator [35] to achieve an ideal phase shifter.

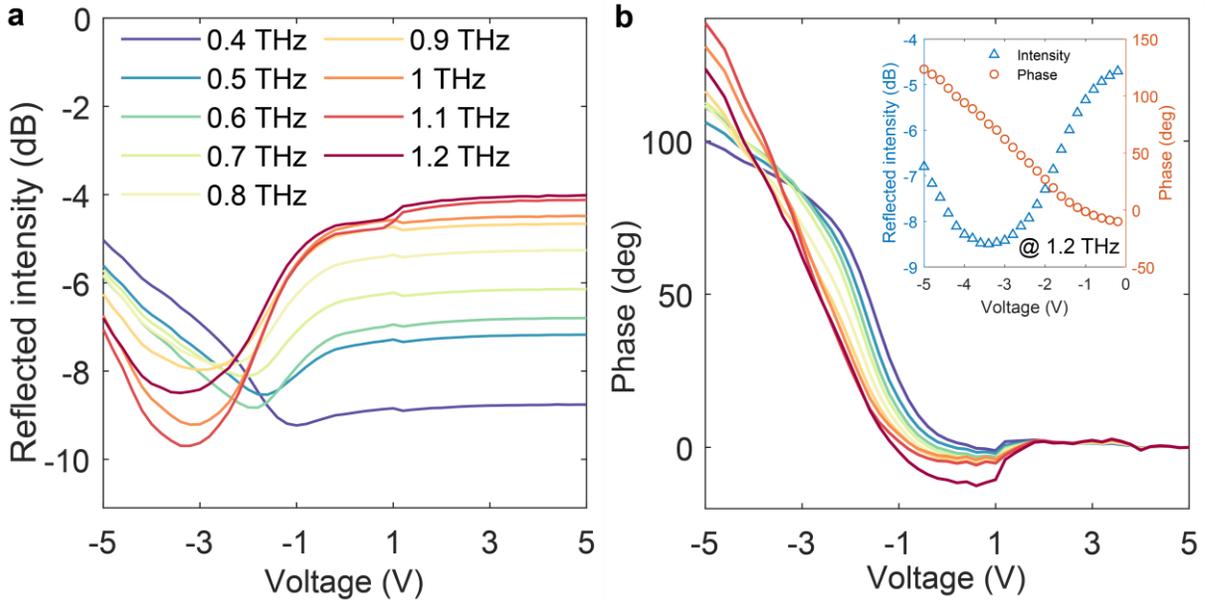

**Fig. 9. The insertion loss and phase shift of D3 under finer gate voltages. a.** Reflected intensity as a function of gate voltage at different frequencies. The reflected intensity is referred the Si prism without device. **b.** Relative phase shift referred to the phase at 5 *V* from 0.4 to 1.2 THz. The inset shows the linearity of the phase shift as a function of driving voltage and the intensity fluctuation.

Modulation speed is another crucial parameter for terahertz modulators. In this study, the modulation speed was evaluated by measuring the direct current (DC) conductivity change of the graphene layer. The THz time-domain system used in this work was not sufficiently fast for direct speed measurements. However, the optical sheet conductivity of graphene is related to its DC conductivity, making it a reasonable proxy for representing the phase modulation speed. For device D3, the measured DC conductivity change speed was approximately 3 kHz (refer to supplementary information for more details), indicating that the phase modulation speed is in the kHz range. To improve the modulation speed of graphene-based devices, one approach is to reduce the resistance-capacitance (RC) time constant. This can be achieved by reducing the size of the graphene or by employing alternative insulation materials. In fact, a graphene modulator operating in the MHz range has been reported by reducing the graphene size and using different insulation materials [36]. It is worth noting that the tunable conductivity interface in the total internal reflection (TIR) geometry, as demonstrated in this work, is not limited to graphene. Other materials and techniques can also be explored for achieving phase modulation. For instance, utilizing other semiconductor systems such as two-dimensional electron gases in heterostructures holds the potential for improving the operation speed to the GHz range [37] in the future.

The current device serves as a proof-of-concept design, and there are several methods to further increase the phase shift. Firstly, by using a different Si prism with a near supercritical incident angle, the phase shift can be enhanced. Secondly, improving the quality of the insulation layer and graphene can expand the conductivity tunable range. Thirdly, optimizing the parameters of the metal grating, such as using finer metal microslits, can lead to a deep subwavelength scale and enable the device to operate over an even wider bandwidth [29]. This optimization can further enhance the phase modulation capabilities. Additionally, an alternative approach is to utilize a Fresnel rhomb made of high-resistivity silicon. The Fresnel rhomb can introduce multiple reflections to the terahertz wave, thereby increasing the overall phase shift up to $2\pi$ with minimal additional loss inside the rhomb. By combining these methods and further optimizing the device design, it is possible to achieve higher phase shifts in future iterations of the terahertz modulator.

## 4. Conclusions

In summary, an ultrawideband terahertz phase shifter based on the TIR geometry was proposed and experimentally demonstrated. The key component of the device was a tunable conductive interface realized by a graphene-loaded non-resonant metasurface. By adjusting the gate voltage, the device achieves continuous and tunable phase shifting of the reflected terahertz waveform. The average phase shift over the frequency range of 0.4 to 1.2 THz is approximately 120°, with minimal amplitude fluctuation of less than 1 dB. The insertion loss of the graphene device, excluding the loss introduced by the Si prism, is below 10 dB, surpassing the performance of metamaterial-based terahertz modulators in terms of bandwidth and insertion loss. The phase modulation speed of the device is in the kHz range, which can be further improved through parameter optimization or by utilizing other semiconductor systems, potentially reaching the GHz range. In the future, there is the possibility of extending the single element to a phase modulator array. This could be valuable for applications such as terahertz ultrawideband phase array radar or imaging systems, where the precise control of phase shifting across multiple elements is required.

**Acknowledgements**

The authors gratefully acknowledge partial financial support for this work from the National Natural Science Foundation of China [grant numbers 61975135, 61805148], International Cooperation and Exchanges NSFC [grant number 61911530218], Shenzhen International Scientific and Technological Cooperation Project [grant number GJHZ20190822095407131] and Natural Science Foundation of Guangdong Province [grant number 2019A1515010869], Guangdong Medical Science and Technology Research Fund [grant number A2020401], Shenzhen University New Researcher Startup Funding [grant number 2019134, RC00058], AoE/P-701/20 by RGC.

# Supplementary information

"Ultrawideband solid-state terahertz phase shifter electrically modulated by tunable conductive interface in total internal reflection geometry"

Xudong Liu[1], Daosong Yu[1], Chuanfu Sun[1], Zhijie Mei[1], Hao Chen[1], Jianbin Xu[2*], Yiwen Sun[1*]

[1] Department of Biomedical Engineering, School of Medicine, Shenzhen University, Shenzhen 518060, China

[2] Department of Electronic Engineering, The Chinese University of Hong Kong, Hong Kong SAR, China

## 1. The derivation of phase equation for conductive interface total internal reflection model

The reflection coefficient of the conductive interface total internal reflection (CI-TIR) geometry with metal microslits can be written as:

$$r_s = \frac{E_i}{E_r} = \frac{n_1 \cos\theta_i - i \cdot \sqrt{n_1^2 \sin^2\theta_i - n_2^2} - \eta \cdot Z_0 \sigma_s}{n_1 \cos\theta_i + i \cdot \sqrt{n_1^2 \sin^2\theta_i - n_2^2} + \eta \cdot Z_0 \sigma_s}. \quad (1)$$

The numerator represents the electric field of incident terahertz (THz) light, while the denominator represents the electric field of reflected terahertz light. The phase of the reflection coefficient corresponds to the phase difference between the incident and reflected components, and it is depicted in the complex plane in Supplementary Figure 1.

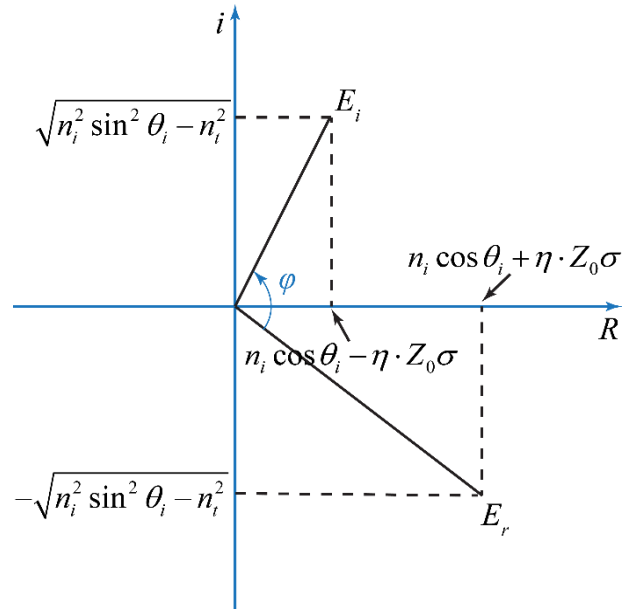

**Supplementary Fig. 1:** The phase of the reflection coefficient in the conductive interface total internal reflection (CI-TIR) geometry with metal microslits is represented in complex plane. $E_i$ is the incident electric field and $E_r$ is the reflected electric field.

Thus, the phase is:

$$\varphi = \tan^{-1}\frac{\sqrt{n_i^2 \sin^2\theta_i - n_t^2}}{n_i \cos\theta_i - \eta \cdot Z_0 \sigma} + \tan^{-1}\frac{\sqrt{n_i^2 \sin^2\theta_i - n_t^2}}{n_i \cos\theta_i + \eta \cdot Z_0 \sigma}. \quad (2)$$

## 2. The phase change of graphene-based Brewster angle device

The reflection coefficient for *p*-polarization light in paper of graphene-based Brewster angle device can be expressed as [1]:

$$r_p = \frac{\sqrt{\varepsilon_s \mu_s}\cos\theta_i - \sqrt{\varepsilon_0 \mu_0}\cos\Phi + Z_0 \sigma_g \cos\theta_i \cos\Phi}{\sqrt{\varepsilon_s \mu_s}\cos\theta_i + \sqrt{\varepsilon_0 \mu_0}\cos\Phi + Z_0 \sigma_g \cos\theta_i \cos\Phi}. \quad (3)$$

The phase has a discontinuous and abrupt change when the sheet conductivity $\sigma_g$ is real. The result is illustrated by Supplementary Figure 2.

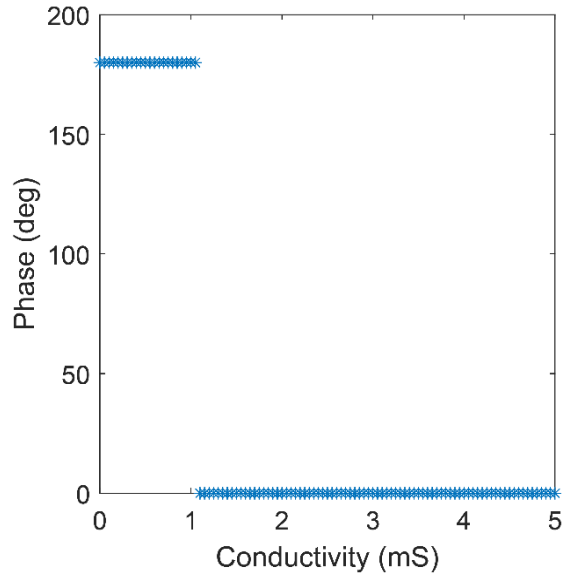

**Supplementary Figure 2:** The phase of the reflection coefficient $r_p$ as a function of $\sigma_g$ according to Eq. 3 in Ref [1]. The incident angle is 68° and the materials are the same as those in Ref [1].

If all the parameters in Eq. 3 are real, including $\sigma_g$, the phase of the reflection coefficient $r_p$ will only exhibit an abrupt jump of 180° when the numerator changes from positive to negative. This behavior is depicted in Supplementary Figure2. However, in Ref. [1], the phase can only be changed by the complex value of $\sigma_g$, which has a limited tunable range. As stated in the supplementary information of Ref [1], at low frequency (e.g., 0.2 to 0.8 THz), the real part of $\sigma_g$ can be tuned from 0.5 to 4, but its imaginary part can only be tuned from $0.01i$ to $1i$. Consequently, the phase modulation of the Brewster angle device at low frequencies can only have an abrupt change. In Fig. 5b of Ref [1], an evident phase jump is observed at low frequencies. Even at frequencies above 0.8 THz, where the imaginary part of graphene becomes larger, the phase modulation still exhibits a jump. As a result, the Brewster angle device does not function as a continuous phase shifter.

**3. The phase change range of liquid crystal material in the TIR geometry**

Due to the birefringence of liquid crystal (LC) materials in the terahertz band, they have been explored as phase shifters in both transmission [2, 3] and the TIR geometry [4]. However, the birefringence of LC in the THz band is typically small, usually less than 0.2 [5], resulting in a limited phase change capability. For instance, the LC material RDP-94990 has $n_e$=1.75 and $n_o$=1.55 in the frequency range of 0.5 to 2.5 THz. Supplementary Figure 3 illustrates the phase change as a function of the incident angle in the TIR geometry for both *s*- and *p*- polarization. It can be observed that the phase difference between the extraordinary (solid line) and ordinary (dashed line) cases is limited.

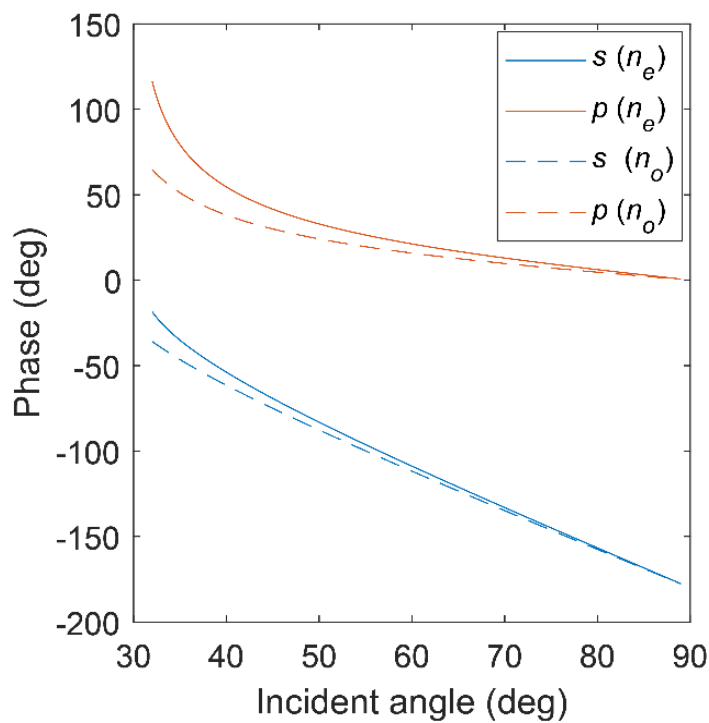

**Supplementary Figure 3:** The calculated phase of reflected light in the TIR geometry for both *s*- and *p*- polarization with RDP-94990.

# 4. Optical sheet conductivity of the graphene in this experiment in the terahertz band

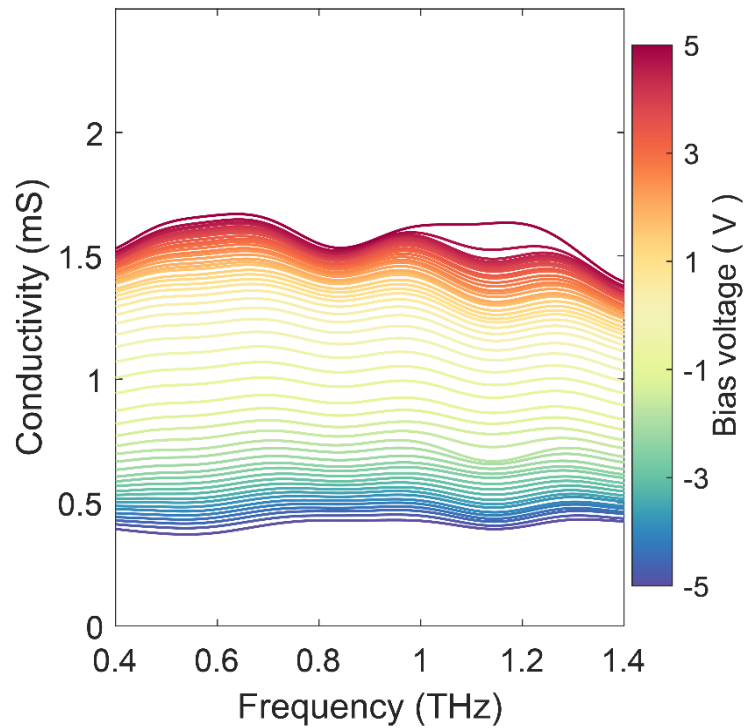

**Supplementary Figure 4:** Optical sheet conductivity of graphene as a function of applied gate voltage in the terahertz band.

In our experiment, the optical sheet conductivity of graphene was measured using a THz time-domain spectroscopy system in the transmission geometry. By applying a gated voltage ranging from –5 to 5 *V*, the optical sheet conductivity of graphene could be tuned within the range of approximately 0.3 to 1.5 *mS*. It was observed that the rate of change in sheet conductivity became slower when the gated voltage was between 1 and 5 *V*.

# 5. Device fabrication

The metallization process of the metal microslits involved the following steps:

1. Deposition of a 5 *nm* thick titanium layer: A 5 *nm* thick layer of titanium was first deposited onto the substrate as an adhesive layer. The titanium layer helps to improve the adhesion between the substrate and the subsequent gold layer.
2. Deposition of a 200 *nm* thick gold layer: After the titanium layer, a 200 *nm* thick layer of gold was deposited on top.
3. Transfer of graphene: Once the metal microslits were fabricated, a piece of graphene was transferred onto the area covering the microslits. The graphene layer was carefully positioned and adhered to the metal surface, creating the conductive interface necessary for the terahertz phase modulation.

This metallization process ensured the presence of a conductive layer (graphene) on top of the metal microslits, enabling the tunability of the conductivity and subsequent phase modulation of the terahertz waves.

Table 1. The measured metal microslit width.

| No.<br>Slit | D2 | D3 | D4 | D5 | D6 | D7 | D8 | D9 | D10 |
|---|---|---|---|---|---|---|---|---|---|
| μm | ~1 | ~2.2 | ~3.3 | ~4.4 | ~5 | ~6.4 | ~7 | ~7.65 | ~9.1 |

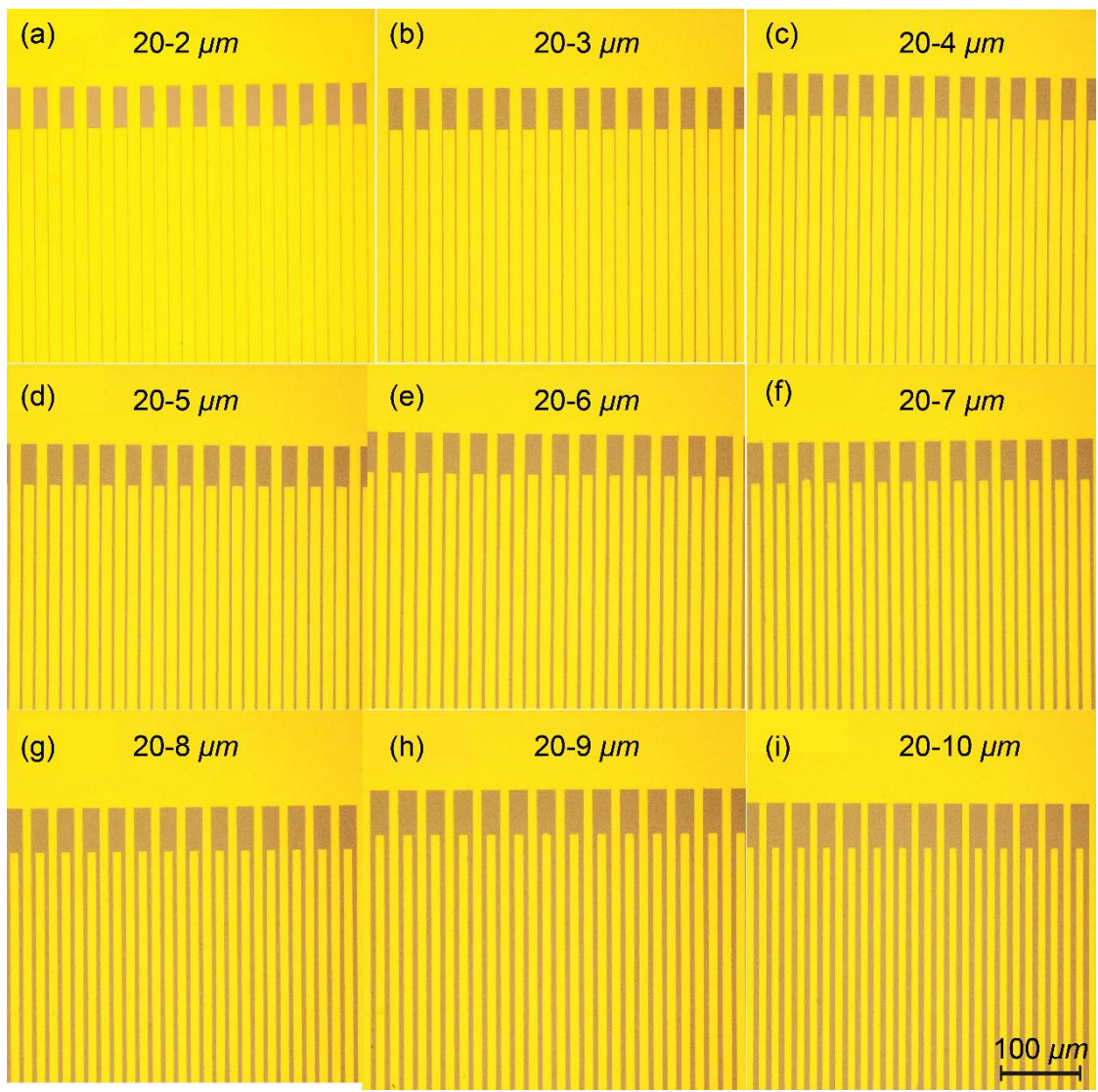

**Supplementary Figure 5:** Optical micrograph of the metal microslits from D2 to D10.

## 6. Graphene characterization

The characteristic Raman spectrum of a representative spot on the graphene of D3, which indicates that the CVD-grown graphene was highly crystalline, as seen from the 2D/G intensity ratio.

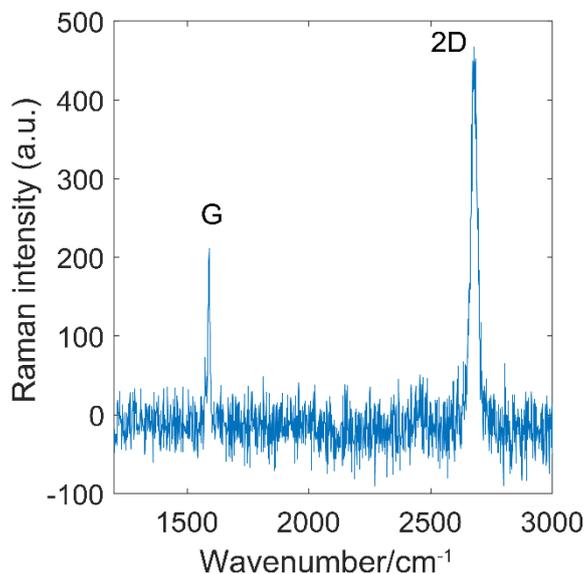

**Supplementary Figure 6:** Representative Raman spectrum of one spot on the

Graphene of D3 measured under 532 *nm* laser excitation.

## 7. The experimental setup

A fibre coupled commercial terahertz time-domain spectroscopy system (TERA-K15, Menlo Systems Inc.) was used to perform the reflection experiment. A homemade optical cage system was used to collimate the THz beam. Two probes were used to apply gate voltage between the graphene and the device substrate.

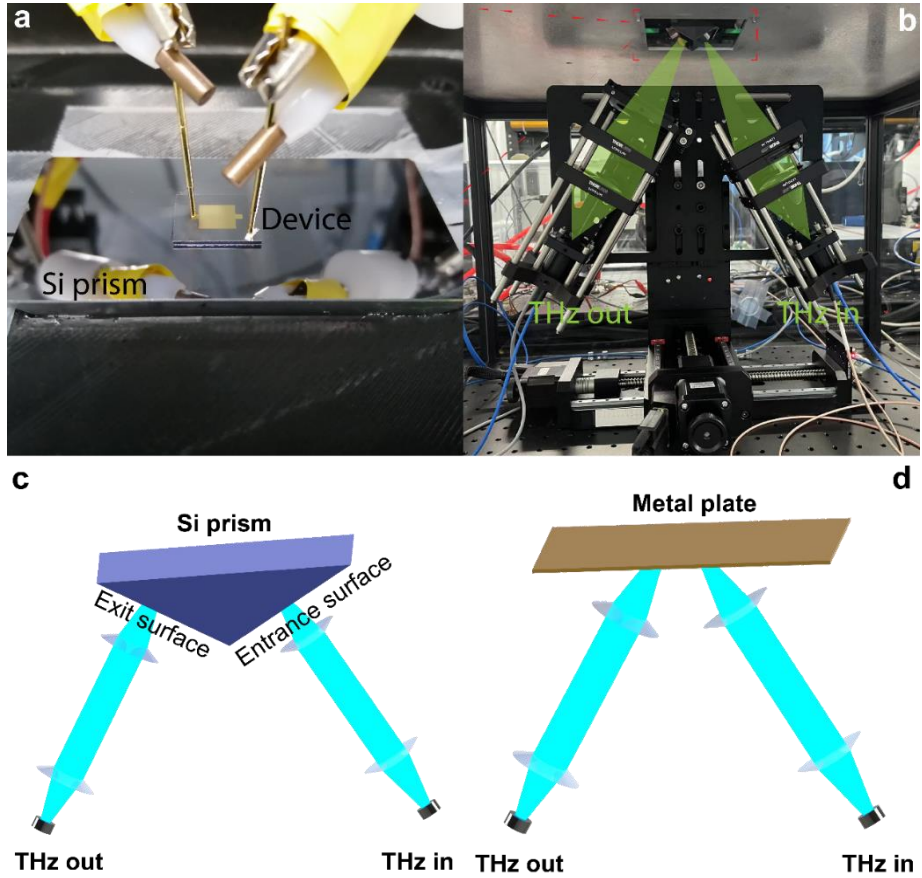

**Supplementary Figure 7:** The experimental setup in this paper. **a.** The high-resistivity silicon prism and the graphene-load metal microslits device. Two probes were used to apply the bias voltage. **b.** The THz time-domain spectroscopy system, which has 30° incident angle to the normal. **c.** with a bare Si prism serves as a reference for calculating the insertion loss of the entrance and exit surfaces. **d.** The experiment using metal plate as a reference to calibrate the electric field amplitude of the THz-TDs system. Measure the intensity of the incident light before it enters the prism. This represents the reference intensity.

## 8. The experiment with finer sweeping voltage step

Supplementary Figure 8a shows the terahertz waveform was gradually delayed as the driving voltage swept from –5 $V$ to 5 $V$ in the time-domain. Supplementary Figure 8b shows the frequency-domain amplitude of a metal plate reference, the Si prism and D3 at various voltages, which was used to calculate the insertion loss.

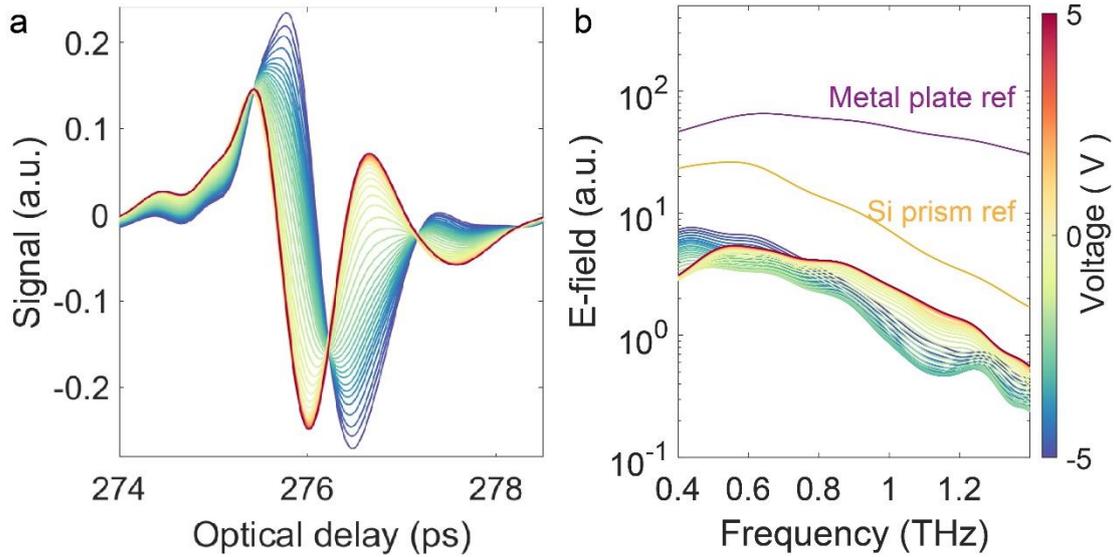

**Supplementary Figure 8: a.** The terahertz waveform in time-domain as a function of driving voltage from –5 to 5 $V$ as 0.2 $V$ as a step. **b.** The frequency-domain amplitude of metal plate reference (purple line), Si prism reference (yellow line) and D3 at voltages from –5 to 5 $V$, which was used to calculate the insertion loss.

## 9. Operation speed measurement

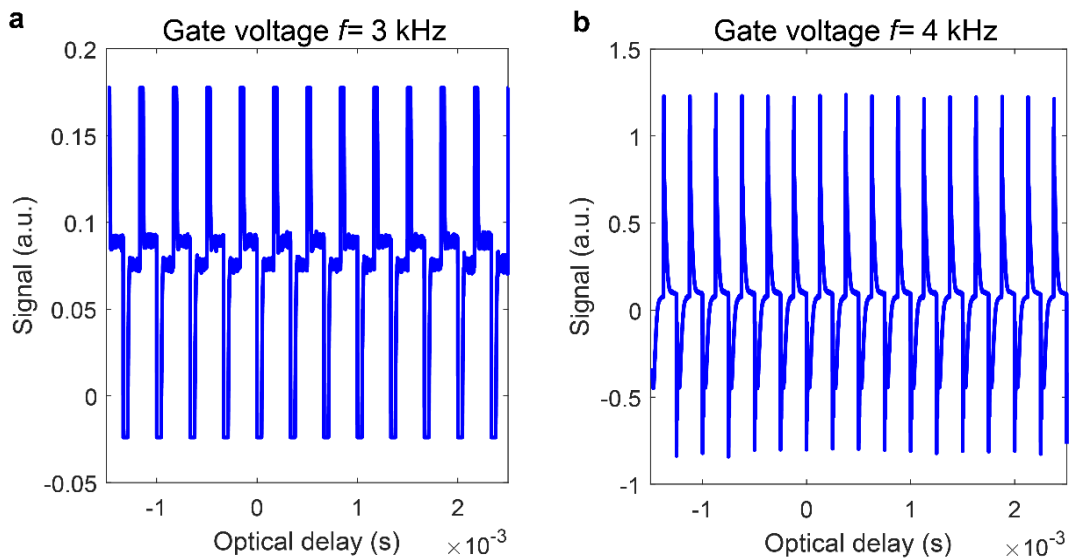

**Supplementary Figure 9:** The drain-source current as a function of gate voltage frequency. **a.** The gate voltage is a quare wave at frequency of 3 kHz. **b.** The gate voltage is a quare wave at frequency of 4 kHz.

The operation frequency of the phase shifter described in the paper is determined by the frequency of the DC conductivity change. This change in conductivity is related to the optical sheet conductivity of graphene. In order to measure the current flowing between the drain and source electrodes, a square wave voltage ranging from –3V to 3V is applied as the gate voltage at a specific frequency. When the gate voltage is applied, the current between the drain and source electrodes is measured. The sharp peaks observed in the measurement correspond to the charging and discharging processes occurring within the graphene device. In Supplementary Figure 9b of the paper, it is shown that the graphene device is not able to achieve a stable state, indicating that it is approaching the operational speed limit. This suggests that the device is reaching the maximum frequency at which it can reliably function.

**Supplementary References:**